\documentclass[acmsmall]{acmart} %
\AtBeginDocument{%
  \providecommand\BibTeX{{%
    \normalfont B\kern-0.5em{\scshape i\kern-0.25em b}\kern-0.8em\TeX}}}

\newcounter{rq}[section]

\newcounter{futureopp}[section]
\newenvironment{futureoppenv}[1][]{\refstepcounter{futureopp}\textbf{Future Opportunity \thefutureopp #1}}{}
\usepackage{pifont}

\usepackage{booktabs}
\usepackage{lscape}
\usepackage{longtable}

\begin{document}

\title{Designing AI Learning Experiences for K-12: Emerging Works, Future Opportunities and a Design Framework}

\author{Xiaofei Zhou}
\affiliation{%
  \institution{University of Rochester}
  \streetaddress{250 Hutchison Rd}
  \city{Rochester}
  \state{NY}
  \country{USA}}
\email{zhouxf626@gmail.com}

\author{Jessica Van Brummelen}
\affiliation{%
  \institution{Massachusetts Institute of Technology}
  \streetaddress{77 Massachusetts Avenue}
  \city{Cambridge}
  \state{MA}
  \country{USA}}
\email{jess@csail.mit.edu}

\author{Phoebe Lin}
\affiliation{%
  \institution{Harvard University}
  \streetaddress{13 Appian Way}
  \city{Cambridge}
  \state{MA}
  \country{USA}}
\email{phoebelin@gsd.harvard.edu}

\begin{abstract}
Artificial intelligence (AI) literacy is a rapidly growing research area and a critical addition to K-12 education. However, support for designing tools and curriculum to teach K-12 AI literacy is still limited. There is a need for additional interdisciplinary human-computer interaction and education research investigating (1) how general AI literacy is currently implemented in learning experiences and (2) what additional guidelines are required to teach AI literacy in specifically K-12 learning contexts. In this paper, we analyze a collection of K-12 AI and education literature  to show how core competencies of AI literacy are applied successfully and organize them into an educator-friendly chart to enable educators to efficiently find appropriate resources for their classrooms. We also identify future opportunities and K-12 specific design guidelines, which we synthesized into a conceptual framework to support researchers, designers, and educators in creating K-12 AI learning experiences.

\end{abstract}



\maketitle

\section{Introduction} %
The rise of artificial intelligence (AI) and machine learning (ML) in user-facing technologies has led to increased interaction between people and AI/ML systems in everyday contexts  \cite{witten2002data}. In particular, children are very likely to interact frequently with AI systems (i.e. Siri and Alexa, or Youtube) in various contexts, such as the home or at school. Despite this increased interaction, children often cannot recognize when they interact with AI, nor how the AI works. This has led to a growing AI literacy gap \cite{how-smart-toys-druga,hey-google-druga}. Consequently, AI and education researchers have called for designing better learning experiences that foster AI literacy \cite{big5} to empower all learners to be critical consumers of AI technology. At the same time, incorporating AI in K-12 education offers young learners the opportunity to see themselves as future builders of AI \cite{a-is-for-ai}.

The work that has been published within the last few years has provided a strong foundation for the growing field of AI literacy. In the last year, Long and Magerko published one of the first frameworks for AI literacy designed to empower non-technical learners to effectively interact and evaluate AI technologies \cite{AILiteracy}. However, most of the research in AI literacy applies to general audiences, and few have outlined design considerations for K-12 learning contexts. Within the K-12 AI education body of work, even fewer are accessible to K-12 educators who typically have limited prior experience with teaching AI \cite{teacher-education,one-year}.

AI education that is designed for a general audience can typically be applied in adult and university-level learning contexts. In contrast, K-12 learning contexts have unique needs (e.g., emphasis on engagement and scaffolding) and challenges (e.g., tight school schedules, student developmental milestones), requiring additional design considerations to overcome them. Therefore, there is a clear need to investigate how the design of AI learning tools and curriculum may address K-12 learning contexts. In this paper, we build on existing work and analyze the methods with which general design considerations and competencies are applied in existing tools and curriculum. Furthermore, we explored specifically how design considerations may differ in K-12 learning contexts and where there are gaps in the literature.

We conducted an exploratory review of AI for K-12 education (AI4K12) literature and tools in order to distill a design framework to inform development of AI learning experiences for K-12. We examined work that falls under AI education and AI literacy, but narrowed our search within a K-12 learning grade band, though we occasionally drew inspiration from adjacent work in other fields or higher learning grade bands (i.e., general computer science education and university-level education). We then organized key design elements in a conceptual framework that we thematically derived from the literature.

The main contributions of this paper are:
\begin{enumerate}
    \item An analysis of how existing work applies general AI literacy competencies and design considerations;
    \item A design framework for designers and researchers creating K-12 AI education tools;
    \item A reference chart for educators selecting AI educational tools to suit the needs of their classrooms.
\end{enumerate}

As K-12 AI education continues to grow and evolve, we expect this analysis and framework to be further refined as well. We present this paper as a stepping stone for other designers and researchers to create better AI educational experiences for K-12 learners, so that their students may also become the designers and researchers of AI education in the future.

\section{Background} %
\citeauthor{big5} pioneered identifying what K-12 students should know about AI. In \citep{big5}, they listed a variety of AI teaching and learning resources for K-12, presented five \emph{Big AI Ideas} for young learners, and elaborated on them for four grade bands: K-2, 3-5. 6-8, and 9-12. Many recent works have used the \emph{Big AI Ideas} to develop effective K-12 AI tools and curriculum \cite{one-year, zhorai, toivonen2020co, ali2019constructionism}.

Due to the nascency of the field, few literature reviews on AI/ML education for K-12 students exist. In one of the most recent reviews, \citeauthor{teaching-ml-systematic-mapping} conducted a systematic mapping with a focus on ML instructional units (IUs) in elementary to high schools. The researchers set their scope with respect to the IUs (``a set of classes [...] designed to teach certain learning objectives to a specific target audience''), presented the ML competencies currently taught in existing IUs, the application domains covered, the types of data utilized, and the types of instructional methods and materials used. They emphasized the importance of teaching young students about AI, how to implement the IUs as extracurricular units, a need for K-12 AI assessment, and the trade-off between creating transparent systems to provide more comprehensive information and simplifying or ``black-boxing'' underlying ML concepts to avoid cognitive overload. They also identified a lack of training for K-12 teachers to implement the IUs in their classrooms. Nonetheless, the authors also mentioned how existing IUs not published in scientific articles (e.g., tools developed by those outside the research community) were not evaluated. Furthermore, they only analyzed ML IUs, rather than AI curricula in general. 

Another study, \cite{AILiteracy}, analyzed curricula with a broader focus on AI. The objective of the study was to create an \emph{AI Literacy} framework for developing tools and curricula for a general, non-technical audience, including learners outside of K-12 \citep{AILiteracy}. The authors thematically derived a list of AI competencies and design considerations from existing research and AI learning resources %
and proposed a conceptual framework to guide HCI researchers' in developing AI curricula. This study focused less on evaluating how existing research has implemented AI literacy elements, but rather on the rationale behind the framework itself. Thus, to address the gaps discussed in this study as well as in \citep{teaching-ml-systematic-mapping}, we evaluate existing K-12 AI education research using the AI literacy framework, point out needs specific to a K-12 audience, identify opportunities for future K-12 AI curricula based on current implementations, and propose a design framework for K-12 AI tools and curricula.

\section{Methodology} %
We conducted an exploratory review of the existing AI learning tools and curriculum in the literature in order to analyze how designs contributed to building K-12 AI literacy. We collected all existing work using %
the ``snowballing'' \cite{wohlin2014guidelines} and keyword-search approaches. We started the snowballing approach from the two most relevant literature reviews on general AI literacy \cite{AILiteracy} and teaching ML in K-12 schools \cite{teaching-ml-systematic-mapping}. To capture the latest work, we collected additional papers from recent conference proceedings including CHI 2020, IDC 2020, IUI 2020, TEI 2020, AIED 2020. Finally, we searched for the terms or a combination of the terms ``Artificial Intelligence'', `Machine Learning'', ``K-12'', ``elementary/primary school'', ``middle school'', ``high school'' on Google Scholar and the ACM library. Since the field is still emerging and growing, and potential AI literacy-adjacent work may not be presented explicitly as such, the list we captured may not be exhaustive. %

We started by reading three representative works: one describing a system to teach K-12 students AI/ML literacy \citep{uncovering-black-boxes-gesture}, one describing a middle school AI curriculum \citep{one-year}, and one outlining design considerations of general AI literacy \citep{AILiteracy}. Three members of our team generated an initial 10 criteria---which was eventually expanded to 14---with which to evaluate subsequent papers. The additions allowed us to evaluate each AI tool and curriculum through the lens of a K-12 learning context. Those three members then met weekly to discuss the criteria and continuously iterate on the evaluation results. This ensured the coding schemes to be consistent and sufficiently comprehensive to generate novel insights to inform future work in the K-12 AI education domain. Partway through the analysis, Cohen's Kappa was calculated for inter-coder reliability for each criteria to ensure consistency across all codes and researchers. The ones with under-satisfactory agreement \cite{lazar2017research} were discussed thoroughly and resolved accordingly.

\section{Overview of Existing Trends}
Recent research has emerged that focuses on introducing AI/ML knowledge to young learners. Several major trends include 1) Structured series courses teaching basic ML concepts and utilizing existing AI education tools \cite{one-year,multiyear-k6,ml-for-all-k-12-vonwangenheim, irobot}; 2) short workshops using interactive data visualization and toy problems to teach students basic algorithms \cite{intro-ml-secondary-essinger,srikant2017introducing, ml-at-high-school}; 3) learning environments enabling students to develop basic AI applications with block-based programming \cite{snap-kahn,ai-appinv,pic-danny,ml-apps-kevin}; 4) Accessible and engaging GUI/TUI/VUIs enabling students to train and test ML models with much of the programming complexity hidden behind the interface \cite{toivonen2020co,smileycluster,uncovering-black-boxes-gesture,tensorflow-playground}.

\section{Current Implementations of AI Literacy Guidelines and Future Opportunities} %

\begin{figure}[htb!]
\caption{List of AI Competencies from \protect{\citealt{AILiteracy}}.}\label{tab:competencies}
\small
\begin{tabular}{ll|ll}
    \# & Competency                   & \# & Competency                   \\ \hline
    1  & Recognizing AI               & 10 & Human Role in AI             \\
    2  & Understanding Intelligence   & 11 & Data Literacy                \\
    3  & Interdisciplinarity          & 12 & Learning from Data           \\
    4  & General vs. Narrow           & 13 & Critically Interpreting      \\
    5  & AI Strengths \& Weaknesses   &    & Data                         \\
    6  & Imagine Future AI            & 14 & Action \& Reaction           \\
    7  & Representations              & 15 & Sensors                      \\
    8  & Decision-Making              & 16 & Ethics                       \\
    9  & ML Steps                     & 17 & Programmability              
\end{tabular}
\end{figure}

This section draws on literature to describe the AI literacy guidelines we used to analyze current implementations of K-12 AI curricula. Specifically, we adapt \citeauthor{AILiteracy}'s AI literacy framework (see Tab. \ref{tab:competencies} and \ref{tab:considerations}) to a K-12 context and add criteria to better address the scope and needs of this context. We also summarize existing implementations %
and how they relate to the additional criteria. Table \ref{tab:educators-table} presents a portion of our analysis of the works with respect to competencies and other metrics relevant to educators (e.g., target age groups).

\begin{figure}[htb!]
\caption{List of AI Design Considerations from \protect{\citealt{AILiteracy}}.}\label{tab:considerations}
\small
\begin{tabular}{ll|ll}
    \# & Consideration                & \# & Consideration                   \\ \hline
    1  & Explainability               & 9 & Identity, Values, \& Backgrounds \\
    2  & Embodied Interactions        & 10 & Support for Parents             \\
    3  & Contextualizing Data         & 11 & Social Interaction              \\
    4  & Promote Transparency         & 12 & Leverage Learners' Interests    \\
    5  & Unveil Gradually             & 13 & Acknowledging Preconceptions    \\
    6  & Opportunities to Program     & 14 & New Perspectives                \\
    7  & Milestones                   & 15 & Low Barrier to Entry            \\
    8  & Critical Thinking            &    &
\end{tabular}
\end{figure}

Furthermore, we indicate which elements of the framework were infrequently addressed in the literature. For example, only 13 of the 49 works explicitly addressed \emph{Ethics} and only six addressed \emph{Natural Interaction}---two of \citeauthor{big5}'s ``Big Ideas''. %
Fewer than ten works addressed each of \emph{Recognizing AI}, \emph{Understanding Intelligence}, \emph{Imagining Future AI}, %
\emph{AI's Strengths \& Weaknesses}, %
and \emph{Action \& Reaction}, and fewer than four works introduced \emph{General vs. Narrow AI} and \emph{Interdisciplinarity}.

In addition to competencies being infrequently addressed, certain design considerations were also overlooked. We noticed polarizing trends, such as how over half of the works \emph{Leveraged Learners' Interests}, created a \emph{Low Barrier to Entry}, \emph{Promoted Transparency}, utilized \emph{Explainability}, \emph{Contextualized Data} and provided students with \emph{Opportunities to Program}, whereas fewer than 15 of the 49 works utilized the design considerations, \emph{Milestones}, \emph{Critical Thinking} and \emph{New Perspectives}, fewer than ten utilized \emph{Embodied Interactions} and \emph{Unveil Gradually} (with respect to the tool, as described below), and no more than three utilized \emph{Identity, Values, \& Background}, \emph{Acknowledging Preconceptions} and \emph{Support for Parents}. This imbalance in addressing the AI literacy framework's elements provides opportunities for future work, which we label as \textbf{Future Opportunities} below.

\subsection{What is AI?}
In the AI literacy framework \citep{AILiteracy}, four competencies describe the answer to ``What is AI?'': \emph{Recognizing AI}, \emph{Understanding Intelligence}, \emph{Interdisciplinarity}, and \emph{General vs. Narrow}. A variety of activities %
addressed this question, including asking students to suggest reasons for why or why not items (e.g., an automatic door) used AI \cite{howtotrainrobot}, introducing students to how social media often uses AI \citep{decoding-ethical-design}, and encouraging discussion about machine vs. human learning \citep{ml-at-high-school}. Other activities included presenting multiple AI technologies using different types of AI, such as knowledge-based, supervised ML and generative AI systems \citep{popbots,irobot}. However, only one work \citep{one-year} directly addressed the concept of \emph{General vs. Narrow AI}; thus, there is an opportunity (\begin{futureoppenv}\label{FO:general-narrow}\end{futureoppenv}) to create curriculum teaching students how much of current AI is narrow in scope, and can only do very specific tasks. Furthermore, only three directly addressed the breadth of types of AI (\emph{Interdisciplinarity}), presenting \begin{futureoppenv}\label{FO:interdiscip}\end{futureoppenv}.

\subsection{What can AI do?}
The competencies, \emph{AI's Strengths \& Weaknesses} and \emph{Imagine Future AI}, address the question ``What can AI do?''. Although many works allowed students to engage with AI systems with specific strengths and weaknesses, only four facilitated discussion about or explicitly indicated AI's capabilities and limitations (\begin{futureoppenv}\label{FO:strengths-weaknesses}\end{futureoppenv}). These works either asked students about the effectiveness of specific types of AI (e.g., clustering, conversational AI) \cite{smileycluster,vanbrummelen-sm} or described and allowed students to engage with AI's ability to both fail and succeed \cite{ml-at-high-school,zhorai}. Less than 15\% of the works addressed \emph{Imagining Future AI}. Those that did engaged students in envisioning activities \citep{one-year,vanbrummelen-sm,cognimates}, discussed how future work or systems would change \citep{alternate-curriculum,kids-making-ai}, or encouraged students to design new AI systems \cite{ai-appinv,vanbrummelen-sm,decoding-ethical-design}.

\subsection{How does AI work?}
There are many AI literacy competencies to consider when answering the question, ``How does AI work?'' Two of the competencies (\emph{Representations} and \emph{Decision-Making}) are associated with cognitive systems; five (\emph{ML Steps}, \emph{Human Role in AI}, \emph{Data Literacy}, \emph{Learning from Data} and \emph{Critically Interpreting Data}) are associated with ML; and two (\emph{Action \& Reaction} and \emph{Sensors}) are associated with robotics. Additionally, we consider \emph{Programmability} to be associated with each system, as it is the underlying mechanism enabling AI to work.

To teach students about cognitive systems, curricula generally discussed representations, like data structures \citep{irobot} or visualizations \citep{zhorai}, and how AI makes decisions \cite{irobot,logic-programming}. Many works also engaged students in developing AI systems and observing how they make decisions based on particular datasets. Since ML systems make decisions---whether about classifications, clustering structures or path sequences---the \emph{ML Steps} competency is also highly related to \emph{Decision-making}.

Many works focused on ML through empowering students to gather data for, train, test and deploy ML systems. This included image classifiers \citep{pic-danny,snap-kahn,ml4kids}, gesture classifiers \citep{sports-ml-zimmermann, youth-gesture-zimmermann, gestures-scratch}, and natural language processing systems \citep{ml4kids}. These curricula engaged students in the process of developing ML systems, teaching them the \emph{ML Steps} as well as about the \emph{Human Role in AI} through experience. Other works taught students about the human role through engaging them in programming AI systems \citep{pedagogical-simulation,popstar-poet-grinch,multiyear-k6,logic-programming}. Many of the curricula that taught students about the \emph{ML Steps} also taught students about \emph{Learning from Data}, \emph{Critically Interpreting Data} and \emph{Contextualizing Data} through encouraging students to collect data and critically analyze how the AI system learns from it. 

These competencies are also highly related to the \emph{Data Literacy} concept, which was taught through exemplifying the importance of quality data \citep{classroom-ai-primary-ho}, data fitting in ML algorithms \cite{ml-at-high-school}, and creating hypotheses and testing them based on experimental data \citep{pedagogical-simulation}, among other methods. One work describing how to embed media literacy in education includes \citealt{media-literacy-ml-valtonen}. It specifically focuses on how data literacy can be taught with respect to specific technologies, ML, and computer science, and provides a call to action for media literacy across educational disciplines.

Curricula we reviewed also taught students about how AI systems can affect the physical world through sensing (\emph{Sensors}) the environment and reacting (\emph{Action \& Reaction}). These works used robots, including the \emph{Cozmo} robot \citep{cozmo}, which sensed and picked up blocks; a toy vehicle \citep{personalizing-homemade-bots}, which sensed a racetrack and followed it; and Arduino robots \citep{howtotrainrobot}, which were programmed in Scratch to sense and react according to students' interests.

\subsection{How should AI be used?}
The \emph{Ethics} competency is at the heart of the question, ``How should AI be used?''. Many works that addressed \emph{Critically Interpreting Data} had additional components in which students discussed the ethics of data collection and ML bias \citep{ml-at-high-school,zhorai,ai-appinv}. Other curricula taught ethics through real-life or system-specific examples, such as the ethics of conversational agents, like Alexa, or robotic agents \citep{vanbrummelen-sm,howtotrainrobot,popstar-poet-grinch}. One curriculum's main focus was ethics. It engaged students in stakeholder analysis, redesigning YouTube with an ethical lens, and recognizing systems as socio-technical \citep{decoding-ethical-design}. This work emphasized how all technology has ethical implications, presenting us with \begin{futureoppenv}\label{FO:ethics}\end{futureoppenv} to encourage researchers to embed ethical discussions and activities into all AI curricula. Another work discussed how specific aspects of ethics can be incorporated into curricula \cite{ai-ethics-equity-edu-southgate}, and may be used as a framework for further AI ethics integration.

\subsection{How can we teach AI?}
The final question we present is, ``How can we teach AI?'', which is addressed by the AI literacy design considerations \citep{AILiteracy} (see Tab. \ref{tab:considerations}). Through reviewing the literature, we found connections between the considerations and group them here with respect to accessibility for learners, engagement, views on AI, and social and developmental considerations.

\subsubsection{Accessibility for Learners}
To teach AI, concepts must be made accessible to learners. Many works we reviewed utilized the \emph{Explainability} design consideration by providing visualizations \citep{youth-gesture-zimmermann,smileycluster,vanbrummelen-sm,ml-at-high-school,zhorai,intro-ml-secondary-essinger}, interactive activities \citep{cozmo,vanbrummelen-sm,pic-danny,youth-gesture-zimmermann} and in-depth explanations \citep{vanbrummelen-sm,youth-gesture-zimmermann,irobot}. Many explainable tools also \emph{Promoted Transparency} through revealing how different training datasets influenced ML decisions \citep{gestures-scratch,ml4kids,ml-apps-kevin,pic-danny,uncovering-black-boxes-gesture}. Other curricula promoted transparency through detailing particular algorithms' steps \citep{intro-ml-secondary-essinger,smileycluster}. Nevertheless, novices can be overwhelmed by seeing too much information at once \citep{teaching-ml-systematic-mapping}, which may occur with transparent systems. Thus, tools and curricula should also consider how to present information to prevent information overload, which may be achieved using the \emph{Unveil Gradually} and \emph{Low Barrier to Entry} considerations. 

We found that although many curricula slowly revealed information through scaffolding \citep{irobot,pic-thesis-danny,multiyear-k6,data-science-summer-mobasher,ai-appinv,ml-for-all-k-12-vonwangenheim}, only five tools innately unveiled information gradually (\begin{futureoppenv}\label{FO:unveil-gradually}\end{futureoppenv}). One such tool included \emph{Zhorai}, which unveiled content gradually through providing a simple natural interaction activity first, then a representation activity, and finally, an activity combining natural interaction, representation and ML \citep{zhorai}. Other works \emph{Lowered the Barrier to Entry} through other methods, including opting for visual, block-based coding \citep{popstar-poet-grinch,vanbrummelen-sm,ai-appinv,ml-apps-kevin,pic-danny,cognimates,gestures-scratch,child-friendly-programming-kahn} instead of text-based coding, and providing approachable visualizations \citep{ml-at-high-school,zhorai,youth-gesture-zimmermann,smileycluster}.

\subsubsection{Engagement}
\label{section: engagement}

The AI literacy framework presents a number of design considerations to promote learner engagement. Many works provided students with \emph{Opportunities to Program} \citep{logic-programming,gender-diversity-ai-summer-vachovsky,alternate-curriculum,ai-appinv,intro-ml-secondary-essinger}, which engaged students in the \emph{Programmability} and \emph{Human Role in AI} competencies. Through self-directed programming projects, many of these works also \emph{Leveraged Learners' Interests} \citep{vanbrummelen-sm,youth-gesture-zimmermann,snap-kahn,gestures-scratch,ml-apps-kevin}. Other works leveraged interests through engaging students in sports or other content outside of computer science and AI \citep{kids-making-ai,sports-ml-zimmermann,zhorai,classroom-ai-primary-ho}. Another promising opportunity (\begin{futureoppenv}\label{FO:identity-values-bg}\end{futureoppenv}) to leverage interests includes engaging students in investigating their \emph{Identity, Values \& Backgrounds}. Only three of 49 works utilized this design consideration. Those that did engaged students with their cultural background in agriculture \citep{kids-making-ai}, addressed the challenges female students faced due to their gender \citep{gender-diversity-ai-summer-vachovsky}, and encouraged students to develop projects with environmental values \citep{ml-for-all-k-12-vonwangenheim}.

Curricula also engaged students through \emph{Embodied Interactions}. For instance, one work included an activity in which students physically acted out feature extraction and the nearest-neighbor algorithm \citep{classroom-ai-primary-ho}; in another, students acted as rule-based conversational agents \citep{vanbrummelen-sm}. Only six works engaged students in embodying full algorithms or processes such as these \citep{classroom-ai-primary-ho,vanbrummelen-sm,ml-children-who-whom-vartiainen,ai-cs-k-uni-kandlhofer,ai-simulation-game-opel,multiyear-k6} (\begin{futureoppenv}\label{FO:embodiment}\end{futureoppenv}). Nonetheless, some works involved shallow embodiment in which students embodied data collection through gestures \citep{gestures-scratch,youth-gesture-zimmermann,sports-ml-zimmermann,uncovering-black-boxes-gesture}. These works \emph{Contextualized Data} through empowering students to observe how their own data could be used to affect AI systems' decisions. Other curricula contextualized data through collecting other types of data, such as images or words \citep{ml4kids,pic-thesis-danny,ml-apps-kevin}, and data relevant to their lives, such as Tweets or course-related data \citep{gender-diversity-ai-summer-vachovsky,ml-at-high-school,intro-ml-secondary-essinger}.

\subsubsection{Views on AI}

Another method of contextualizing is to \emph{Acknowledge Preconceptions} about AI, such as how AI is portrayed in the media. Only two of the 49 works we reviewed alluded to acknowledging preconceptions (\begin{futureoppenv}\label{FO:acknowledge-preconceptions}\end{futureoppenv}): one observed how students perceived AI to be highly related to robotics and computers \citep{ml-at-high-school}, and another asked students to identify values of potential AI stakeholders, enabling researchers to address preconceptions in the values \citep{decoding-ethical-design}. Both of these works went on to discuss the reality of the AI field, encouraging \emph{New Perspectives} on AI. Other works broadened students' understanding of AI through presenting less-popular methods of AI, like logic-based learning \citep{irobot}, or engaging students in activities emphasizing different decision-making techniques \citep{popbots,cognimates}. Many of these works also engaged students in \emph{Critical Thinking} through discussing preconceptions and how they might change. Other works encouraged critical thought through ethical discussion, touching on the \emph{Ethics} competency and often the \emph{Critically Analyzing Data} competency.

\subsubsection{Social and Developmental Considerations}
\label{section: parents}

Fourteen of the 49 works considered students' developmental stage to better attune curricula to their cognitive abilities, addressing the \emph{Milestones} consideration. Some works did so by referring to developmental psychology research or educational standards  \citep{any-cubes-toy-scheidt,logic-programming,intro-ml-secondary-essinger,multiyear-k6,smileycluster,popbots}, and others did so through working with teachers who had experience engaging with students of specific age levels \citep{one-year,ml-for-all-k-12-vonwangenheim}. \emph{Cognimates}, a curriculum for younger students, additionally considered how students may interact with their parents, and developed a learning guide for them \citep{cognimates}. Only three other works involved parents, engaging them in the AI activity, if there were not enough students \citep{zhorai}, inviting them to the students' final presentations \citep{data-science-summer-mobasher}, or having them informally teach their children concepts \citep{ml-children-who-whom-vartiainen}. Future work might consider how to better involve parents in the learning experience (\begin{futureoppenv}\label{FO:involve-parents}\end{futureoppenv}), possibly involving them in a co-design process or encouraging student-parent collaborative learning. Although most works did not involve parents, many involved peers, addressing the \emph{Social Interaction} design consideration. Some curricula included collaborative group activities or discussions \citep{zhorai,coming-to-senses-lechelt,ml-for-all-k-12-vonwangenheim,one-year}; others included group projects \citep{vanbrummelen-sm,gestures-scratch,pic-thesis-danny,data-science-summer-mobasher}; and others involved group competitions or games \cite{kids-making-ai,classroom-ai-primary-ho}.

\begin{landscape}
\footnotesize
\begin{longtable}{@{}l|cccc|cc|ccccccccc|cc|llllllll@{}}
\caption{Our evaluation of 49 K-12 AI education works based on \protect{\citeauthor{AILiteracy}}'s AI literacy framework. It can also serve as a quick-reference for educators to implement learning activities based on AI literacy competencies, target age groups (Targ.), time required for the activity, physicality of the tool (i.e., if it requires a physical robot, a computer, etc.), %
amount of scaffolding (Scfld.), %
low- or high-ceilingness of the activity (Ceil.), which refers to how the activity can be extended beyond the provided curriculum, and amount of teacher involvement when developing and implementing the tool (Teach.). The table is sorted by competency (i.e., first sorted by Competency 1, then 2, etc.). \protect\footnotemark}
\label{tab:educators-table}\\
 & \multicolumn{17}{c|}{Competencies} &  &  &  &  &  &  &  &  \\
 & \multicolumn{4}{c}{What is AI?} & \multicolumn{2}{c|}{\begin{tabular}[c]{@{}c@{}}What can \\ AI do?\end{tabular}} & \multicolumn{9}{c|}{How does AI work?} & \multicolumn{2}{c|}{\begin{tabular}[c]{@{}c@{}}How should\\ AI be used? \&\\ programmability\end{tabular}} &  &  &  &  &  &  \\
Key & 1 & 2 & 3 & 4 & 5 & 6 & 7 & 8 & 9 & 10 & 11 & 12 & 13 & 14 & 15 & 16 & 17 & Targ. & Time & Tool & Scfld. & Ceil. & Teach. \\* \midrule
\endhead
\footnotetext{Targ.: P/M/H/U/A/N - Primary/Middle/High/Undergrad/All/Not specified; \\
Time: H/D/M/Y/N - measured in Hours/Days/Months/Years/Not specified; \\
Tool requirement: P/W/U - Physical(tangible)/Web(software)/Unplugged (Note: Physical trumped Web trumped Unplugged); \\
Scaffolding: I/T/N - Instructional scaffolding / Tool scaffolding / Not specified; \\
Ceilingness: H/L - High/Low ceiling; \\
Teacher involvement: R:I/T:I/T:L/T:D/N - Researchers as Instructors / Teachers as Instructors/ Teachers as Learners / Teachers as Designers / Not specified}

One-Year \citep{one-year} & \ding{51} & \ding{51} &  & \ding{51} &  & \ding{51} & \ding{51} & \ding{51} & \ding{51} & \ding{51} &  & \ding{51} & \ding{51} &  & \ding{51} & \ding{51} & \ding{51} & M & Y & W & I & H & T:D \\
Robot \citep{howtotrainrobot} & \ding{51} & \ding{51} &  &  &  &  &  &  &  & \ding{51} &  &  &  & \ding{51} &  & \ding{51} & \ding{51} & M & W & P & I & H & R:I \\
Cognimate \citep{cognimates} & \ding{51} &  &  &  &  & \ding{51} & \ding{51} & \ding{51} & \ding{51} & \ding{51} &  &  &  &  & \ding{51} &  & \ding{51} & P & N & W & I & H & R:I \\
Alternate \citep{alternate-curriculum} & \ding{51} &  &  &  &  & \ding{51} &  & \ding{51} & \ding{51} &  &  & \ding{51} &  &  & \ding{51} & \ding{51} &  & H & W & W & I & L & T:I \\
Decoding \citep{decoding-ethical-design} & \ding{51} &  &  &  &  & \ding{51} &  &  & \ding{51} &  &  & \ding{51} &  &  &  & \ding{51} &  & M & H & U & I & L & N \\
For-All \citep{ml-for-all-k-12-vonwangenheim} & \ding{51} &  &  &  &  &  & \ding{51} & \ding{51} & \ding{51} & \ding{51} & \ding{51} & \ding{51} & \ding{51} &  & \ding{51} & \ding{51} & \ding{51} & M & M & W & I & H & T:I \\
K-Uni \citep{ai-cs-k-uni-kandlhofer} & \ding{51} &  &  &  &  &  & \ding{51} & \ding{51} & \ding{51} &  & \ding{51} & \ding{51} &  &  &  &  &  & A & W & P & I & H & R:I \\
Base \citep{shamir2020transformations} & \ding{51} &  &  &  &  &  &  & \ding{51} & \ding{51} & \ding{51} & \ding{51} & \ding{51} &  &  &  &  & \ding{51} & P & N & W & T & L & N \\
Co-design \citep{toivonen2020co} & \ding{51} &  &  &  &  &  &  & \ding{51} & \ding{51} & \ding{51} &  & \ding{51} & \ding{51} &  &  &  & \ding{51} & P & D & W & I & H & R:I \\
IRobot \citep{irobot} &  & \ding{51} & \ding{51} &  &  &  & \ding{51} & \ding{51} &  &  &  &  &  & \ding{51} &  &  & \ding{51} & H & W & P & I & H & R:I \\
Classroom \citep{classroom-ai-primary-ho} &  & \ding{51} & \ding{51} &  &  &  &  & \ding{51} & \ding{51} & \ding{51} & \ding{51} & \ding{51} &  &  &  &  & \ding{51} & M & W & P & I & L & T:I \\
Conver \citep{vanbrummelen-sm} &  & \ding{51} &  &  & \ding{51} & \ding{51} & \ding{51} & \ding{51} &  & \ding{51} &  & \ding{51} &  &  &  & \ding{51} & \ding{51} & M/H & W & W & T & H & R:I \\
Why-Not \citep{ml-at-high-school} &  & \ding{51} &  &  & \ding{51} &  & \ding{51} & \ding{51} & \ding{51} & \ding{51} & \ding{51} & \ding{51} & \ding{51} &  &  & \ding{51} &  & H & D & W & T & L & N \\
Game \citep{ai-simulation-game-opel} &  & \ding{51} &  &  &  &  &  & \ding{51} & \ding{51} &  &  & \ding{51} &  &  &  &  &  & M/H & H & U & I & N & T:L \\
Popbots \citep{popbots} &  &  & \ding{51} &  &  &  & \ding{51} & \ding{51} & \ding{51} & \ding{51} &  & \ding{51} &  & \ding{51} &  &  & \ding{51} & P & W & P & I & L & N \\
Smiley \citep{smileycluster} &  &  &  &  & \ding{51} &  & \ding{51} & \ding{51} & \ding{51} & \ding{51} & \ding{51} & \ding{51} &  &  &  &  &  & H & H & W & T & L & N \\
Zhorai \citep{zhorai} &  &  &  &  & \ding{51} &  & \ding{51} & \ding{51} & \ding{51} & \ding{51} &  & \ding{51} & \ding{51} &  &  & \ding{51} &  & P & H & W & T/I & L & R:I \\
STEM \citep{kids-making-ai} &  &  &  &  &  & \ding{51} & \ding{51} & \ding{51} & \ding{51} & \ding{51} & \ding{51} & \ding{51} & \ding{51} &  &  &  & \ding{51} & M & D & W & I & H & T:I \\
App-Inv \citep{ai-appinv} &  &  &  &  &  & \ding{51} & \ding{51} & \ding{51} & \ding{51} & \ding{51} &  & \ding{51} & \ding{51} &  & \ding{51} & \ding{51} & \ding{51} & M/H & W & W & I & H & N \\
Wolfram \citep{wolfram2017elementary} &  &  &  &  &  &  & \ding{51} & \ding{51} & \ding{51} & \ding{51} & \ding{51} & \ding{51} &  &  &  &  &  & A & N & W & I & H & N \\
Summer \citep{data-science-summer-mobasher} &  &  &  &  &  &  & \ding{51} & \ding{51} & \ding{51} & \ding{51} & \ding{51} & \ding{51} &  &  &  &  &  & H & W & W & I & H & T:D \\
Tensorflow \citep{tensorflow-playground} &  &  &  &  &  &  & \ding{51} & \ding{51} & \ding{51} & \ding{51} &  & \ding{51} & \ding{51} &  &  &  &  & H/U & N & W & N & H & N \\
Sports \citep{sports-ml-zimmermann} &  &  &  &  &  &  & \ding{51} & \ding{51} & \ding{51} & \ding{51} &  & \ding{51} & \ding{51} &  &  &  &  & M/H & H & P & I & H & R:I \\
Gentle \citep{gentle-scratch-estevez} &  &  &  &  &  &  & \ding{51} & \ding{51} & \ding{51} & \ding{51} &  & \ding{51} &  &  &  &  & \ding{51} & H & W & W & I & H & T:I \\
High-Sch \citep{mariescu2019machine} &  &  &  &  &  &  & \ding{51} & \ding{51} & \ding{51} & \ding{51} &  & \ding{51} &  &  &  &  &  & P/H & H & W & I & H & T:I \\
Youth \citep{youth-gesture-zimmermann} &  &  &  &  &  &  & \ding{51} & \ding{51} & \ding{51} &  &  & \ding{51} & \ding{51} &  & \ding{51} &  & \ding{51} & P/M & D & P & I & H & R:I \\
Nodes \citep{gestures-scratch} &  &  &  &  &  &  & \ding{51} & \ding{51} & \ding{51} &  &  & \ding{51} & \ding{51} &  &  &  & \ding{51} & N & N & P & N & H & N \\
Control \citep{pedagogical-simulation} &  &  &  &  &  &  & \ding{51} & \ding{51} &  & \ding{51} & \ding{51} &  &  &  & \ding{51} &  & \ding{51} & U & N & W & N & H & N \\
Bots \citep{personalizing-homemade-bots} &  &  &  &  &  &  & \ding{51} & \ding{51} &  & \ding{51} &  & \ding{51} & \ding{51} & \ding{51} &  &  &  & N & N & P & N & L & N \\
Cozmo \citep{cozmo} &  &  &  &  &  &  & \ding{51} & \ding{51} &  & \ding{51} &  &  &  & \ding{51} & \ding{51} &  & \ding{51} & A & N & P & I & H & N \\
Logic \citep{logic-programming} &  &  &  &  &  &  & \ding{51} & \ding{51} &  & \ding{51} &  &  &  &  &  &  &  & M & W & W & I & H & T:D \\
Semantic\citep{ketamo2009semantic} &  &  &  &  &  &  & \ding{51} & \ding{51} &  &  &  &  &  &  &  &  &  & P & H & W & T & L & N \\
Secondary \citep{intro-ml-secondary-essinger} &  &  &  &  &  &  & \ding{51} &  & \ding{51} & \ding{51} &  & \ding{51} & \ding{51} &  &  &  &  & H & N & W & N & H & T:I \\
Grinch \citep{popstar-poet-grinch} &  &  &  &  &  &  & \ding{51} &  &  & \ding{51} & \ding{51} & \ding{51} &  &  &  & \ding{51} & \ding{51} & N & N & W & N & H & N \\
Gender \citep{gender-diversity-ai-summer-vachovsky} &  &  &  &  &  &  & \ding{51} &  &  & \ding{51} & \ding{51} &  & \ding{51} &  &  &  &  & H & W & W & I & H & T:I \\
Boxes \citep{uncovering-black-boxes-gesture} &  &  &  &  &  &  &  & \ding{51} & \ding{51} & \ding{51} &  & \ding{51} & \ding{51} &  &  & \ding{51} &  & M & N & P & I & L & N \\
Image \citep{pic-thesis-danny} &  &  &  &  &  &  &  & \ding{51} & \ding{51} & \ding{51} &  & \ding{51} & \ding{51} &  &  &  & \ding{51} & H & D & W & I & H & T:I \\
Apps \citep{ml-apps-kevin} &  &  &  &  &  &  &  & \ding{51} & \ding{51} & \ding{51} &  & \ding{51} & \ding{51} &  &  &  & \ding{51} & H/U & W & W & I & H & N \\
ML4Kids \citep{ml4kids} &  &  &  &  &  &  &  & \ding{51} & \ding{51} & \ding{51} &  & \ding{51} & \ding{51} &  &  &  & \ding{51} & N & N & W & N & H & N \\
Whom \citep{ml-children-who-whom-vartiainen} &  &  &  &  &  &  &  & \ding{51} & \ding{51} & \ding{51} &  & \ding{51} & \ding{51} &  &  &  & \ding{51} & P & H & W & I & H & R:I \\
Snap! \citep{snap-kahn} &  &  &  &  &  &  &  &  & \ding{51} & \ding{51} &  & \ding{51} &  &  & \ding{51} &  & \ding{51} & H & N & W & N & H & N \\
Cloud \citep{child-friendly-programming-kahn} &  &  &  &  &  &  &  &  &  & \ding{51} &  & \ding{51} &  &  & \ding{51} &  & \ding{51} & M/U & N & W & N & H & N \\
Multiyear \citep{multiyear-k6} &  &  &  &  &  &  &  &  &  & \ding{51} &  &  &  &  & \ding{51} &  & \ding{51} & P & Y & P & I & L & T:I \\
Data-Kids \citep{srikant2017introducing} &  &  &  &  &  &  &  &  &  &  & \ding{51} &  &  &  &  & \ding{51} &  & M & H & W & I & L & T:I \\
Any-Cubes \citep{any-cubes-toy-scheidt} &  &  &  &  &  &  &  &  &  &  &  & \ding{51} &  &  &  &  &  & N & N & P & N & H & N \\
Sensors \citep{coming-to-senses-lechelt} &  &  &  &  &  &  &  &  &  &  &  &  &  &  & \ding{51} &  &  & P & H & P & I & H & T:I \\
Cards \citep{bilstrup2020staging} &  &  &  &  &  &  &  &  &  &  &  &  &  &  &  & \ding{51} &  & H & W & U & I & H & T:I \\
Software \citep{sperling2012integrating} &  &  &  &  &  &  &  &  &  &  &  &  &  &  &  &  &  & H & M & W & N & H & T:I
\end{longtable}
\end{landscape}

\section{Extended Design Framework FOR K-12 AI EDUCATION} %

While \citeauthor{big5}'s Big AI Ideas framework and \citeauthor{AILiteracy}'s AI literacy framework have laid the foundation for K-12 AI education guidelines and general AI literacy, there are opportunities to improve K-12 AI curricula within these frameworks. By integrating the future opportunities presented in the previous section (see Tab. \ref{tab:future-opportunities}) with design guidelines from existing AI literacy and adjacent frameworks \cite{AILiteracy} and guidelines \cite{druga2019inclusive}, we propose an extended framework to guide the design of AI learning experiences for K-12 students.

\begin{table}[]
\caption{Future opportunities to strengthen AI tools' and curricula's effectiveness.}\label{tab:future-opportunities}
\small
\begin{tabular}{p{.004\textwidth}p{.28\textwidth}|p{.004\textwidth}p{.28\textwidth}|p{.004\textwidth}p{.28\textwidth}}
\# & Future Opportunity               & \# & Future Opportunity                      & \# & Future Opportunity                         \\ \hline
1  & Explaining General vs. Narrow AI & 4  & Multiple Touchpoints for Ethics         & 7  & Algorithmic Embodiment                     \\
2  & Breadth of Interdisciplinarity   & 5  & Unveil Tools Gradually                  & 8  & Acknowledge Sensationalized Preconceptions \\
3  & Specific Strengths \& Weaknesses & 6  & Address Identity, Values \& Backgrounds & 9  & Support \& Codesign with Parents          
\end{tabular}
\end{table}

\subsection{More Comprehensive AI Literacy as Learning Objectives}
The \textbf{first additional guideline} is related to addressing the first four future opportunities in more depth for a K-12 audience. Future research efforts could introduce more comprehensive definitions of AI (i.e., \textbf{Future Opportunity \ref{FO:general-narrow}, \ref{FO:interdiscip}}), be more explicit about AI's capabilities and limitations (i.e., \textbf{Future Opportunity \ref{FO:strengths-weaknesses}}), and embed ethical discussion about all the aspects of AI technologies (i.e., \textbf{Future Opportunity \ref{FO:ethics}}) \cite{decoding-ethical-design}. \citeauthor{big5}'s Big AI Idea \#4 ``Making agents interact comfortably with humans is a substantial challenge for AI developers'' is so far under-investigated by researchers and educators.

\subsection{Student Engagement}
Engagement is a one of the keys to improve learning outcomes, improve throughput rates and retention, and ensure equality \cite{trowler2010student}. Through our analysis, we summarized additional methods to engage students in learning AI.

\subsubsection{Using Data}
\label{section:engage-data}
Students can find it challenging to make sense of abstract concepts and engage in hypodeductive reasoning practices at different stages of cognitive development \citep{wellman2001meta}. The \textbf{second guideline} has to do with using using concrete data instead of abstract concepts. This can play an essential role in making AI knowledge more accessible for K-12 students, while also engaging them in data collection and data contextualization processes. By creating their own datasets, students can develop Information and Communication Technology (ICT) skills \cite{kaloti2015students}. Such strategies have been implemented in AI education for children (e.g. \cite{mariescu2019machine, uncovering-black-boxes-gesture, sports-ml-zimmermann, any-cubes-toy-scheidt}) and should be investigated further.  %

Using data related to students' social and cultural contexts can also promote students' engagement. For example, \citeauthor{classroom-ai-primary-ho}'s AI learning activities utilized data related to Disney princesses. Other relevant objects and concepts that engaged students in the analyzed works included how animals see, the concept of a ``selfie'', toys \cite{pedagogical-simulation, pic-danny, pic-thesis-danny, any-cubes-toy-scheidt}, and current events (e.g., fake news, latest robots) \citep{decoding-ethical-design,vanbrummelen-sm}. Additionally, such AI learning activities can also tackle real-world problems \cite{ml-for-all-k-12-vonwangenheim, kids-making-ai}, potentially addressing \textbf{Future Opportunity \ref{FO:identity-values-bg}}: \emph{Identity, Values \& Backgrounds}.

\subsubsection{Learn by Teaching}
\citeauthor{druga2019inclusive}'s inclusive AI literacy design guidelines for kids suggested providing children with different ways to teach machines. The \textbf{third guideline} suggests using this learning by teaching method to engage students. Considering the embodied nature of child-computer interaction \cite{papert1990children,ml-children-who-whom-vartiainen}, learning by teaching can be an effective mechanism from constructivist learning theories \cite{kafai1991learning, ackermann2004constructing}. This has also been shown in practice to engage students in AI learning \cite{ml-children-who-whom-vartiainen, zhorai}. In addition to learning through teaching AI agents, students can learn through teaching peers \cite{sperling2012integrating, ml-children-who-whom-vartiainen} (which additionally utilizes the \emph{Social Interaction} design consideration \citep{AILiteracy}). Another method for peer teaching includes ``Jigsaw Learning'' \cite{aronson1978jigsaw, mengduo2010jigsaw}, which was used in a ML activity by \citeauthor{ml-children-who-whom-vartiainen}.

\subsubsection{Gamification and Embodiment}
The \textbf{fourth guideline} was inspired by existing AI education tools that utilize gamification to engage K-12 students. We found that the most common game used for learning was ``Rock Paper Scissors'' \cite{snap-kahn,mariescu2019machine,popbots,teacher-education,one-year,ml-apps-kevin}. %
A number of new educational games have also been designed for teaching AI to K-12 students. For example, the ``AI for Oceans'' \cite{shamir2020transformations} and ``Man, Machine!'' \citeauthor{ai-simulation-game-opel} activities gamify learning ML concepts. In ``Man, Machine!'', players role-play either the humans or machines %
incorporating \textbf{Future Opportunity \ref{FO:embodiment}}: embodiment. Such perspective taking %
is also recommended in the inclusive AI design guidelines for kids \cite{druga2019inclusive}.

\subsection{Built-in Scaffolding}
\subsubsection{Bridge the Gap in Development}
One of the main barriers for K-12 students to learn AI is not having a computer science background. The \textbf{fifth guideline} addresses this barrier. For example, several strategies to overcome this challenge in existing works include 1) using block-based visual programming to train and test ML models \cite{snap-kahn, child-friendly-programming-kahn, gestures-scratch, cognimates}; 2) unveil the complex ML concepts gradually through GUI/TUI/VUIs and visualizations for students to engage with and develop AI systems without programming \cite{uncovering-black-boxes-gesture,smileycluster,intro-ml-secondary-essinger,zhorai} (\textbf{Future Opportunity \ref{FO:unveil-gradually}}); 3) facilitating the progression of AI learning with starter codes, worked examples \cite{gray2007suggestions} or following the ``Use-Modify-Create'' cycle  \cite{ml-for-all-k-12-vonwangenheim, youth-gesture-zimmermann, toivonen2020co, gender-diversity-ai-summer-vachovsky}; and 4) providing detailed workbook or guided tutorials  \cite{logic-programming, ml-apps-kevin,pic-thesis-danny,ai-appinv}.

\subsubsection{Support High Iterativeness with Instant Feedback}
\label{section: iterativeness}
Iteration is an important AI concept. The \textbf{sixth guideline} acknowledges how iteration can also be used to engage children in experimental learning through trial and error \cite{penner1997building}. %
For example, many existing works facilitate quick iteration of machine training and testing and/or interactive feedback \cite{popbots,youth-gesture-zimmermann, gestures-scratch, uncovering-black-boxes-gesture, sports-ml-zimmermann,classroom-ai-primary-ho,howtotrainrobot}, addressing the need to  ``provide meaningful feedback to children with each action they take so they see what the machine has learned or not'' from  \citeauthor{druga2019inclusive}'s inclusive AI literacy design guidelines for kids. Furthermore, utilizing iterativeness to allow AI to make mistakes can strengthen children's understanding of AI's strengths and weaknesses (\textbf{Future Opportunity \ref{FO:strengths-weaknesses}}) and how AI works \cite{zhorai,mariescu2019machine}. 

\subsubsection{Promote Reflection}
Facilitating student reflection can be essential to K-12 AI learning experiences---especially to inspiring new perspectives, engaging in ethics, and acknowledging AI preconceptions (\textbf{Future Opportunity \ref{FO:acknowledge-preconceptions}}). The \textbf{seventh guideline} encourages designers to include additional, diverse reflection activities. In-time feedback loops, as discussed in \ref{section: iterativeness}, data labeling and evaluation \cite{uncovering-black-boxes-gesture}, and peer sharing and remixing \cite{druga2019inclusive} can stimulate student reflection on underlying ML mechanisms. Other methods include after-lab discussion \cite{intro-ml-secondary-essinger, ml-at-high-school,intro-ml-secondary-essinger,coming-to-senses-lechelt,ai-appinv}, reflective worksheets, journals or blogs \cite{alternate-curriculum,coming-to-senses-lechelt,sports-ml-zimmermann,srikant2017introducing,decoding-ethical-design,ml-for-all-k-12-vonwangenheim,ai-simulation-game-opel}, and collaborative group work \cite{kids-making-ai, coming-to-senses-lechelt,bilstrup2020staging}.

\subsection{Teacher and Parent Involvement}
\subsubsection{Teachers-as-Learners, Teachers-as-Instructors, Teachers-as-Designers}
Through education programs in India, one work indicates a need to transform K-12 teachers' self-efficacy and perspective on AI from AI consumers to AI educators \cite{teacher-education}. 
However, another work shows a lack of accessible ML materials and training for K-12 teachers \cite{teaching-ml-systematic-mapping}. We further analyzed teachers' involvement in existing AI learning experiences for K-12 students and found only a few efforts have been made to involve K-12 teachers as instructors, and even fewer to involve them as AI curricula designers. The \textbf{eighth guideline} encourages further involvement with teachers. In \citeauthor{howtotrainrobot}'s work, an accessible curriculum and 2-day teachers-as-learners workshop were designed to equip teachers as AI instructors. In \citealt{ai-simulation-game-opel}, teachers receive a booklet outlining how to facilitate reflective discussion on how ML works. 
In other works, researchers co-designed AI learning experiences with K-12 teachers. For example, to integrate logic programming as a problem-solving tool in K-12 science education, \citeauthor{logic-programming} received feedback from teachers on content appropriateness. %
In \citeauthor{one-year}'s work, ICT teachers were regularly involved in the curriculum development. 

\subsubsection{Design for Parents}
As mentioned in section \ref{section: engagement}, few works designed specific elements for parents in the AI learning experience for K-12 students (\textbf{Future Opportunity \ref{FO:involve-parents}}). One work studied parents' perception %
of a coding toolkit designed for kids and proposed design implications including ``Supporting Parents as Scaffolders'', ``Adding Engaging Features'' and ``Designing for Sibling Play'' \cite{yu2020considering}. \citeauthor{cognimates} designed learning guides for parents, %
encouraging them to learn with their children. The \textbf{ninth guideline} encourages further investigation of what roles parents can play in children's AI learning experience and what can be designed to support parents.

\subsection{Equity, Diversity, and Inclusion}

While many existing works utilize self-generated data in the data collection stage (e.g. selfies, athletic moves) to leverage learners' interest and provide embodied interaction  \cite{sports-ml-zimmermann}, emerging research shows that examining personal data can lead to higher engagement and be effective in incorporating \emph{Identity, Values \& Backgrounds} (\textbf{Future Opportunity \ref{FO:identity-values-bg}}). %
\citeauthor{register2020learning} show novices pay more attention to ML mechanisms when using personal data  and were more able to ground their self-advocacy arguments in the mechanisms \cite{register2020learning}. Though it was not K-12 specific, this work sheds light on the importance of learners' self-awareness in how AI technologies personally affect them. %

More critically, few works addressed how to create equitable, diverse, and inclusive AI learning experiences. One example of encouraging diversity in the field of computer science includes \citeauthor{gender-diversity-ai-summer-vachovsky}'s AI summer camp for high school girls. In this work, researchers helped female students find role models %
and a sense of community in AI computing. The \textbf{tenth guideline} encourages AI education to also be designed to be more equitable, diverse and inclusive. Long-term engagement with marginalized communities (e.g., women, BIPOC, LGBTQ+, and others) requires upfront investment that would move us toward a more diverse AI community. Similarly, we design equitable AI learning experiences for K-12 by investing in appropriate supports for diverse learners, including students with learning disabilities or students from low socioeconomic backgrounds. Inclusive learning environments build safe spaces for learners to practice empathy and give each other grace for mistakes. AI curriculum can introduce scaffolding in ethics and data discussions to promote cross-contextual learning.

\subsection{Integrated AI in Core Curricula}

Due to tight school schedules, it is not always feasible to make space for new K-12 AI curriculum. The \textbf{eleventh guideline} encourages researchers to design at the intersection of AI and other K-12 disciplines to ease integration into the classroom. AI education can contribute to many K-12 curriculum subjects. 
By considering AI as a problem-solving tool, AI learning experience can aim to teach students to apply AI/ML to solve problems under different disciplines, such as science, social studies, (e.g. \cite{ml-for-all-k-12-vonwangenheim,kids-making-ai}). By considering AI as a teachable agent, "learning by teaching" can be applied to enhance learning across various domains. For example, young kids can learn and socialize different emotions while teaching machines to recognize them (e.g. \cite{ml-children-who-whom-vartiainen}). And lastly, by considering AI (specifically ML) as an instance of science practices aligned with K-12 science, engineering, and technology standards \cite{ngsslead2013}, students can develop computational thinking and practice science skills by building ML models (e.g. \cite{sports-ml-zimmermann,smileycluster, popbots}). Non-programming activities can also be designed to practice social studies, science, and math skills (e.g. \cite{howtotrainrobot, ai-simulation-game-opel}). As AI becomes more relevant to different disciplines, curricula that can enable more touch points for interdisciplinary learning will emerge. We hope this design guideline can serve as a starting point to inspire more creative way for K-12 AI education in the future.

\section{Conclusions and Future Work}
This paper provides an analysis of how AI literacy guidelines have been applied in K-12 contexts and nine future opportunities through a survey of AI and education literature. Moreover, we identified eleven additional design guidelines for creating AI learning experiences in K-12 learning contexts, including those related to student engagement, built-in scaffolding, teacher and parent involvement, equity diversity and inclusion, and integrated AI/core curricula. Due to challenges in K-12 education, it is important to discuss the successes and gaps in implementation to inform the future development of learning tools and curriculum. We encourage designers and researchers of K-12 AI education to use these design guidelines to build richer AI learning experiences, and continue refining them as both the field and the needs of the classroom evolve. 
\begin{acks}
\end{acks}

\bibliographystyle{ACM-Reference-Format}
\bibliography{biblio}

\end{document}